\documentclass[sigconf]{acmart}

\copyrightyear{2024}
\acmYear{2024}
\setcopyright{rightsretained}
\acmConference[IoT 2024]{14th International Conference on the Internet of Things}{November 19--22, 2024}{Oulu, Finland}
\acmBooktitle{14th International Conference on the Internet of Things (IoT 2024), November 19--22, 2024, Oulu, Finland}
\acmPrice{}
\acmDOI{10.1145/3703790.3703795}
\acmISBN{979-8-4007-1285-2/24/11}

\usepackage{soul}
\usepackage{algorithmic}
\usepackage{graphicx}
\usepackage{textcomp}
\usepackage{xcolor}
\usepackage{caption}
\usepackage{subcaption}
\usepackage{url}
\usepackage{multirow}
\usepackage{array}
\usepackage{multicol}
\usepackage[inline]{enumitem}
\usepackage{caption}

\begin{document}

\newif\ifhideold
\hideoldtrue 

\newcommand{\comm}[1]{}
\newcommand{\mycomment}[2]{\textcolor{blue}{\textit{#1}}\comm{#2}}
\newcommand{\myNote}[1]{ \textcolor{red}{\textit{Note : #1}} }
\newcommand{\keepOrNot}[1]{\textcolor{gray}{\textit{#1}}}
\newcommand{\hideorshowwork}[1]{
\ifhideold
  
\else
  #1
\fi
}
\newcommand{\takeaway}[1]{
\vspace{2pt}
\noindent
        \textit{\textbf{Takeaway}: #1}
}

\newcommand{\myTodo}[1]{\textcolor{green}{TODO: \textit{#1}}}
\newcommand{\sideNote}[1]{\textcolor{gray}{\textit{#1}}}

\newcommand{\wip}[1]{\textcolor{purple}{wip: \textit{#1}}}
\newcommand{\added}[1]{r: #1}

\title{Towards Adaptive Asynchronous Federated Learning for Human Activity Recognition}

\author{Rastko Gajanin}
\affiliation{\small Distributed Systems Group, TU Wien
    \country{Austria}
}
\email{gajaninrastko@gmail.com}

\author{Anastasiya Danilenka}
\affiliation{\small Faculty of Mathematics and Information Science, Warsaw University of Technology
    \country{Poland}
}
\email{anastasiya.danilenka.dokt@pw.edu.pl}

\author{Andrea Morichetta}
\affiliation{\small Distributed Systems Group, TU Wien
    \country{Austria}
}
\email{a.morichetta@dsg.tuwien.ac.at}

\author{Stefan Nastic}
\affiliation{\small Distributed Systems Group, TU Wien
    \country{Austria}
}
\email{s.nastic@dsg.tuwien.ac.at}

\begin{abstract}


In this work, we tackle the problem of performing multi-label classification in the case of extremely heterogeneous data and with decentralized Machine Learning. Solving this issue is very important in IoT scenarios, where data coming from various sources, collected by heterogeneous devices, serve the learning of a distributed ML model through Federated Learning (FL).
Specifically, we focus on the combination of FL applied to Human Activity Recognition (HAR), where the task is to detect which kind of movements or actions individuals perform. In this case, transitioning from centralized learning (CL) to federated learning is non-trivial as HAR displays heterogeneity in action and devices, leading to significant skews in label and feature distributions. 
We address this scenario by presenting concrete solutions and tools for transitioning from centralized to FL for non-IID scenarios, outlining the main design decisions that need to be taken. 
Leveraging an open-sourced HAR dataset, we experimentally evaluate the effects that \textit{data augmentation, scaling, optimizer, learning rate, and batch size choices} have on the performance of resulting machine learning models. Some of our main findings include using SGD-m as an optimizer, global feature scaling across clients, and persistent feature skew in the presence of heterogeneous HAR data. 
%
Finally, we provide an open-source extension of the Flower framework that enables asynchronous FL.  
\end{abstract}

\maketitle

\sloppy

\section{Introduction}


During the last decade, IoT has been able to connect millions of devices thereby supplying an unprecedented amount of data used by applications to provide innovative services~\cite{maresch2024vate}. Considering individual use, smartphones, smart wristbands, and smartwatches have become part of people's everyday lives. One prominent task is collecting real-time data from individuals to provide assistance; a notable application is human activity recognition (HAR). The HAR rapidly became pivotal in many areas, such as healthcare, human-computer interaction, surveillance systems, entertainment, and more~\cite{s23042182}. The HAR tasks span from recognizing simple common activities, such as walking or running, to assisting in more complex ones, such as doing laundry or preparing meals~\cite{Kumar2024}. Gathering data from various sensors, such as accelerometers, gyroscopes, and magnetometers, becomes essential to precisely model the observed activities. 
At the same time, the pervasiveness of such applications comes with certain costs. It increases the devices' computation demands as well as the need for real-time feedback and it amplifies privacy concerns, especially in sensitive applications~\cite{Saha2024, nastic2022serverlessFabric}.


\textbf{State of the art and beyond.} Federated learning (FL) \cite{mcmahan2017communication} has been emerging with the promise to address these challenges, offering methods for privacy, security, and scalability by decentralizing the machine learning (ML) model training to the clients' devices. 
These characteristics make FL particularly suitable for IoT systems~\cite{9773116}. 
However, dynamic scenarios such as the HAR bring new challenges. The
HAR typically shows broad data heterogeneity due to many individuals performing the same actions differently and with varying frequencies. This scenario produces high intra-class variability and inter-class dissimilarity~\cite{Saleem2023}, making client model training non-trivial. Indeed, it is established that having non-independently and non-identically distributed (non-IID) data across FL clients leads to divergent local model updates and, consequently, undermines global model performance~\cite{https://doi.org/10.48550/arxiv.1806.00582}. 
Furthermore, device heterogeneity in HAR-IoT applications impacts the FL model training by introducing unreliable device connectivity and limitations in storage and computation capabilities~\cite{10492865}. 
To mitigate the challenges associated with real-time updates in such unstable scenarios, asynchronous FL (AFL) was proposed~\cite{xie2019asynchronous}. This approach allows clients to train models and send updates once ready or connected while the server continuously accommodates received model updates. Unfortunately, AFL amplifies the problem of diverging models with non-IID data. In fact, AFL natively favors clients that train faster and communicate more frequently with the server~\cite{XU2023100595}. Therefore, AFL on non-IID data further increases the global variance of model updates~\cite{wang2024tackling}.
As of today, the challenge of non-IID data in asynchronous FL remains open~\cite{XU2023100595}, urging the development of robust methods to address it. 

\textbf{Contributions.} The open challenges in both the HAR and FL research make transitioning from centralized learning (CL) to FL and AFL with non-IID data non-trivial, largely limiting the adoption of FL in practice.
In this paper, we select one of the prominent and robust HAR datasets, \textit{Extrasensory} \cite{vaizman2017recognizing}, for the experimentation and identification of the main challenges inherent to the transition from CL to (A)FL with severe non-IID data. 
Based on this experimentation, we 
define a replicable methodology shaped to guide this transition by focusing on the HAR use-case and non-IID data. We offer key insights on how certain decisions impact AFL and model training and performance in this set of problems. Notably, we implement a publicly available extension of the Flower \cite{beutel2020flower} framework that enables AFL and use it for our evaluation\footnote{Our work is part of the Centaurus Linux Foundation project that provides a novel open-source platform for building unified and highly scalable public or private distributed Edge, Cloud, and 3D continuum systems.}.
In summary:
\begin{enumerate*}[label=\textbf{(\arabic*)}]
    \item We present a realistic HAR case study, where we evaluate the effects of synchronous and asynchronous federated learning and its main optimization strategies in this paradigmatic scenario. This study empirically demonstrates how different design decisions impact model performance, offering key takeaways as a blueprint for similar datasets and tasks in IoT.
    \item We develop a novel, publicly available framework that extends the open-source Flower~\cite{beutel2020flower}, based on the works of FedAsync \cite{xie2019asynchronous}, ASO-Fed \cite{chen2020asynchronous} and PAFLM \cite{lu2020privacy}. We offer practical tools not only to reproduce our results but to implement AFL solutions in other use cases.
    \item We introduce a methodology, which is specifically designed to enable structured transitioning from centralized learning to synchronous and/or asynchronous federated learning in the presence of non-IID data, considering the main design decisions and potential pitfalls. 



    
\end{enumerate*}

\section{Methodology} 
\label{section:methodology}

The general process of developing a centralized ML model is quite standardized and can be summarised (with various levels of detail) into 4 main steps: (1) data analysis and preprocessing
, (2) model training and tuning
, (3) performance evaluation and (4) model deployment. 
When developing federated models with non-IID data, the process becomes non-trivial. This section introduces a methodology that applies common ML design steps while systematically analyzing case-specific decisions for FL systems, revealing their impact on performance in IoT environments.

\subsection{Data Analysis \& Preprocessing}

In CL, full access to the dataset enables the use of sophisticated techniques for data analysis and preprocessing. However, to properly federate these processes, it is necessary to establish the expected level of data heterogeneity and willingness to share local data statistics, potentially introducing additional privacy risks. Thus, one can choose from two main approaches for data analysis and preprocessing in (non-IID) FL:


\begin{enumerate}
    \item Local approach – preprocessing local data based solely on local dataset statistics, for instance, scaling data with local mean and standard deviation or performing data augmentation considering local data label imbalance statistics. 
    \item Global approach –  considering aggregation of data statistics from client devices on the server to form global statistics, which are then shared among all clients, and used to guide local preprocessing.
\end{enumerate}

We illustrate the effects local and global analytics approaches have on scaling and on the resulting model performance for the HAR use case in Section~\ref{subsec:scaling}. We further federate data augmentation strategies in Section~\ref{subsubsec:data_augmentation} and show their effect in FL in Section~\ref{subsec:data_augmentation}.

\subsection{Model Training \& Tuning}
\label{subsec:model_training_tuning}

In CL, the model is trained on a single device, which has consistent access to the entire training dataset. In contrast, in FL training occurs on multiple devices and involves recurrent broadcasting of the global model to devices and aggregating updates from them. Introducing FL affects several design decisions:
\begin{itemize}
    \item The choice of the \textit{model architecture and hyperparameters}, such as batch size and learning rate, due to the data and resource heterogeneity across HAR devices, is influenced by memory constraints, convergence speed, overfitting tracking, and more. We talk more about the effects of hyperparameters in Section~\ref{subsec:hyperparameters_tuning} and Section~\ref{subsec:adam_vs_sgd}
    \item Setting the \textit{number of local epochs} is a new hyperparameter introduced in the FL setting to control the iteration count over the local dataset. Intuitively, it controls to what extent local updates can be fitted toward their own data distributions before synchronizing with the server.
    \item Coordinating the FL training process \textit{synchronously} or \textit{asynchronously}. This decision depends on the reliability and heterogeneity of devices participating in FL, with the latter being more suitable for failure-prone scenarios such as IoT. Depending on whether synchronous or asynchronous FL is chosen, the process of broadcasting and aggregating the model differs, introducing more hyperparameters for training to consider.
\end{itemize}

In \textbf{synchronous} FL the step of model broadcasting involves a set of clients being chosen for the current round of training, parametrized by $S$ - number of clients to choose from the available pool of clients - and a specific client selection strategy (random by default). The server sends them the current global model $x_t$ and instructs them to train the passed global model further with their local data. After local training, the clients send their updated models ($x_{t+1}^i$ for client $i$) back to the server where the server aggregates the resulting client models and updates the global model with this aggregate. Equation~\ref{eq:rule-update-sync} describes the synchronous update procedure ($N$ is the total number of samples and $n_i$ is the number of samples present on client $i$):

\begin{equation}
\label{eq:rule-update-sync}
x_{t+1} = x_{t} +  \sum_{i \in S} \frac{n_i}{N} (x_{t+1}^i)    
\end{equation}

    In \textbf{asynchronous} FL the process of broadcasting and aggregating model updates does not proceed in rounds; contrary to that, clients start training as soon as they merge their newest update with the global model and receive the new merged global model from the server~\cite{xie2019asynchronous,lu2020privacy}. Thus, the model aggregation step needs to account for individual clients' model updates in the current global model. This step is parametrized by \textit{mixing ratio} which dictates the averaging weights (e.g. 50-50 or 30-70) between the global model and the client update. The mixing ratio is multiplied by the proportion of samples held by the client that sends the update ($\frac{n_i}{N}$). 

To formally express the update rule of the asynchronous federated baseline we provide the following Equation~\ref{eq:async-update-rule}. Assume that $\Delta_i$ denotes client $i$'s gradients sent to the server for aggregation. $\alpha_{MR}$ is the above-defined mixing ratio.

\begin{equation}
\label{eq:async-update-rule}
x_{t+1} = x_{t} + \alpha_{MR} \frac{n_i}{N} (x_t + \Delta_i)  
\end{equation}


\subsection{Performance Evaluation \& Model Testing}

In CL, due to the centrally located dataset, partitioning of data can be done using random shuffling and performance evaluation can further be reliably tracked throughout the training process. However, the absence of a centralized dataset affects the formation of train/validation/test sets in FL, allowing for three partitioning scenarios: 
\begin{itemize}
    \item Dividing \textit{clients} into training, validation and test subsets
    \item Dividing \textit{local datasets} and aggregating the reported performance metrics on the server
    \item Applying \textit{hybrid (semi-centralized) techniques}, such as maintaining a test dataset on the server while validating model performance on clients subset or parts of local datasets
\end{itemize}

Without detailed data distribution statistics, ensuring representative data partitions is challenging, leading to skewed performance indicators and suboptimal model performance, further emphasizing the importance of the proper performance evaluation scheme. More details on the implications of the test scheme in non-IID FL are provided in Section~\ref{subsec:test_scheme}.

\subsection{Model Deployment}

Model deployment may be considered the only step in the pipeline where FL has some advantage over CL as in order to deploy the model one should only ensure proper final model broadcasting to all clients, allowing them to start using it for inference.

To sum up, although we can roughly apply the classical CL steps in FL, this distributed setting complicates their proper execution, requiring more ingenuity, especially when facing non-IID data. 

\section{Asynchronous FL with Flower}

To evaluate the challenges of non-IID FL with HAR we selected the Flower framework \cite{beutel2020flower} as the base for our experiments because it is one of the established frameworks for developing and researching FL workflows. However, Flower does not support AFL, which is why we extend the framework to implement it, following the description provided in Section \ref{subsec:model_training_tuning}. We provide our implementation as open-source software to enable quick prototyping and facilitate further research in this area. In the following, we briefly describe the main modifications that enabled AFL with Flower.

We modify the thread pool executor to avoid waiting for clients to finish training. Instead, a callback is triggered upon arrival of each client update and the global model is updated according to Equation~\ref{eq:async-update-rule}. If the maximum train duration is not exceeded it re-submits the client for training with the updated global model. This adaptation is transparent for the participating clients. The server still contains one loop that periodically executes central evaluation until the maximum training duration is exceeded.
Our implementation of AFL in Flower and further implementation details are available in this GitHub repository\footnote{\url{https://github.com/r-gg/flower-async-fork}}.

\section{Case Study: Federated Learning for Human Activity Recognition} 
\label{section:usecase}
In this section we first describe the used HAR dataset and its characteristics. Later, we present the main data engineering adaptations to prepare the data for our approach.

\subsection{Extrasensory Dataset}
Our work leverages the \textit{Extrasensory} dataset \cite{vaizman2017recognizing}. This source contains sensor readings from smartphones belonging to 60 different individuals including such sensors as accelerometer, gyroscope, audio, etc. Time-series-related signal features encompassing various statistical and spectral properties were already extracted for the dataset. The original dataset contains in total 225 features from 11 sensors (sources). As labels, the authors presented 6 primary mutually exclusive labels that describe the individual's current status: \textit{standing, walking, sitting, laying down, running, and cycling}. The individuals themselves reported the labels through a dedicated smartphone app, during or immediately before starting an action/changing status. 
In addition to this set, there exists an expanded set of labels (non-mutually exclusive) that encompass user actions and locations. 
This dataset is ideal for our work because it reflects the real-world IoT setting. The data was collected in-the-wild, thus guaranteeing natural heterogeneities among different clients: individuals have different devices (sensor heterogeneity), different behavior (means of performing actions or being in a certain body state), different habits (certain individuals tend to run more, while others cycle more etc.). The non-IID property of data is simulated in most federated learning research~\cite{xu2023asynchronous}, whereas we use a real case. Furthermore, only a few authors~\cite{shen2022federated, chen2020federated, chen2020asynchronous} tested FL in the Extrasensory dataset and without a detailed analysis.

\subsubsection{Dataset Characteristics}

\begin{figure}
    \centering
    \includegraphics[width=0.8\linewidth]{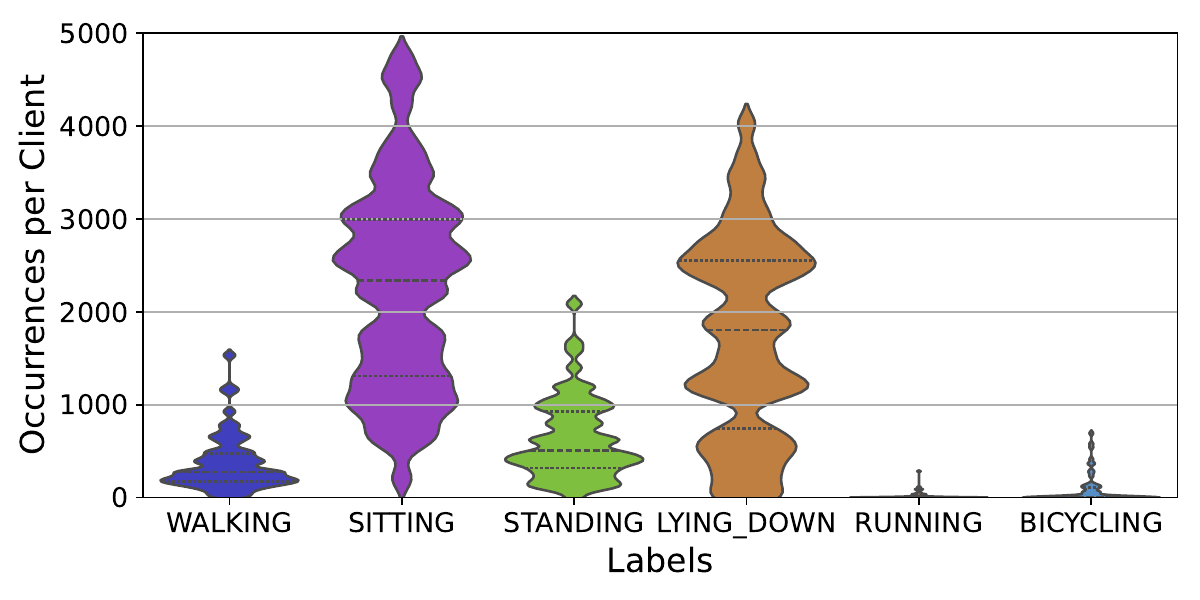}
    \captionsetup{skip=-2pt}
    \caption{Violin plot highlighting the data and label skew. We can see how certain activities are underrepresented, as well as, from the violin \textit{width}, how each activity is not equally distributed across clients.}
    \label{fig:violin-label-no-augmentation}
\end{figure}

The dataset illustrates different types of skew among the clients, namely, label, quantity, and feature skews~\cite{zhu2021federated}. 
Figure~\ref{fig:violin-label-no-augmentation} depicts the data and label skew through a violin plot. The $x$ axis reports each activity, while the $y$ axis shows how many occurrences of that activity are present in each client. The width represents the commonness of having $y$ occurrences for that activity in a client. E.g., if 20 clients have roughly 3000 sitting events, then for the sitting violin plot, at y = 3000 a large width would be present. This plot allows us to see how certain activities are underrepresented, especially running and bicycling (violin height), as well as, how each activity is unequally distributed across clients (violin width). These label and data quantity skews are further complicated by the natural feature skew that is present in the HAR datasets~\cite{Saleem2023} due to individual behavioral patterns expressed by humans, leading to the variability of data representing the \textit{same activity} as well as making similar activities even harder to distinguish. This combination of several data skews materialized in a real-life IoT scenario makes the Extrasensory Dataset a challenging yet valuable candidate for testing solutions for heterogeneous FL.




\subsection{Data Preprocessing}

\noindent \textbf{Data partitioning}
As the data was originally partitioned by individuals we proceeded with this predefined split among the clients. Each client's local dataset was split into three parts following the 64-16-20 \% shuffled split among train-validation and test sets (applying the 80-20 rule consecutively). Each client's test set was sent to the server to create a \textit{fair} centralized test set for evaluation. In the CL setting the local train/validation sets were merged into central train and validation sets.

\noindent \textbf{Feature selection}
For our experiments, we focused on the original set of 6 labels. Data from the majority of the sensors was used as inputs. We discarded the sensors with more than 60\% of missing values in any feature belonging to the respective sensor. The final sensors used for our models are: accelerometer (26 features),  gyroscope (26), watch accelerometer (46), watch compass (9), audio (26), audio properties (2) and phone state as one hot encoded discrete measurements (32), resulting in 175 input features in total. This subset of sensors was selected to avoid handling the many missing values from other sensors and making the input dimension too large. 

\noindent \textbf{Standardization \& Cleaning}
In our preprocessing pipeline, we apply \textit{global standardization}, which sends feature means and standard deviations of all clients to the server in order to create a globally scaled view of the data. We impute the missing values with the feature statistical means.

\subsubsection{Data Augmentation}
\label{subsubsec:data_augmentation}

\begin{figure}
    \centering
    \includegraphics[width=1\linewidth]{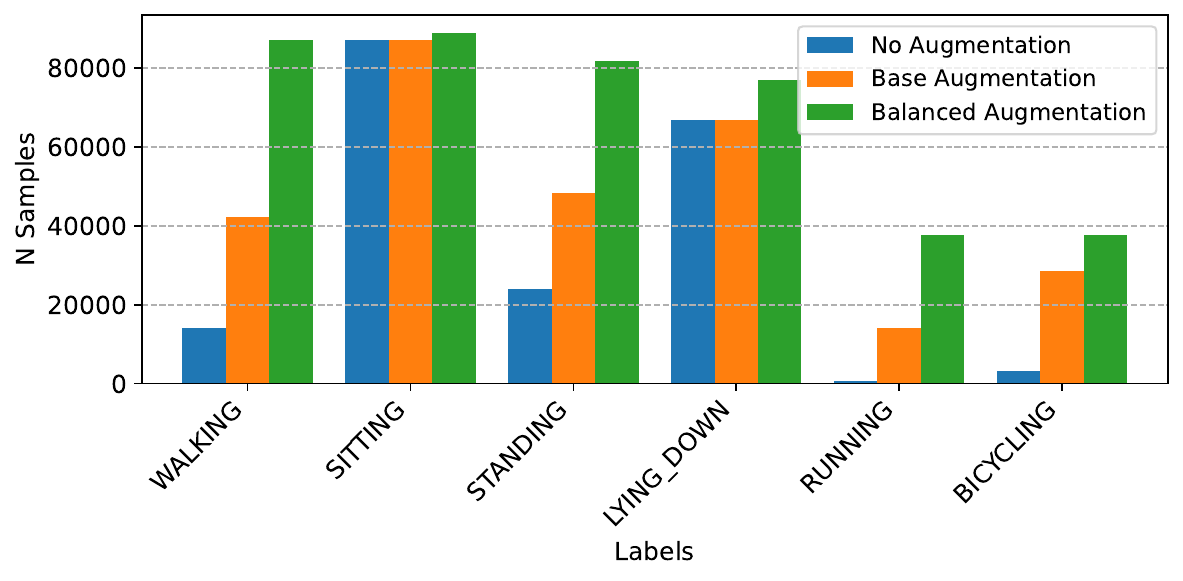}
    \captionsetup{skip=-2pt}
    \caption{Label distribution with different data augmentation settings: \textit{none, base} and \textit{balanced}. Note that in the balanced setting \textit{running} and \textit{cycling} are still \textit{less represented} globally.}
    \label{fig:label_distribution_different_augmentation}
\end{figure}

As the classes are already severely imbalanced without any augmentation (See Figure \ref{fig:label_distribution_different_augmentation}), we perform data augmentation \textit{after} standardization and missing value imputation. We extract and replicate samples of each class an empirically-defined number of times (Running 20 times, Cycling 8 times, Standing 1 time, Walking 2 times). Then, we use Gaussian noise to augment the features of each replica (Mean 0 and std $10^{-4}$). We call this augmentation stage \textit{``base''}. We also examine the \textit{balanced} augmentation setting; here, all existing labels are balanced \textit{on each client separately}. In this case, the number of replicas separately created for each sample dictates the balance augmentation result. This number is the ratio of the number of samples of the most common label over the number of samples of the currently augmented label. (e.g. if sitting is the most common label on the client $i$ and has $n_s^i$ samples, then the number of replicas of the running samples on the same client will be $\lfloor \frac{n_s^i}{n_r^i} \rfloor$.  The resulting label distributions are depicted in Figure \ref{fig:label_distribution_different_augmentation}, which shows how, previously negligible labels, like \textit{running} and \textit{bicycling} have a better representation and better compensate for the data skew. In particular, the balanced augmentation brings labels like \textit{waling} and \textit{standing} to have almost the same volume as the prominent one, i.e., \textit{sitting}.

\subsection{Quality Assessment Metrics}
\label{subsec:qam}
Another essential step is to define a clear set of quantitative measures of the model's quality. In particular, we want metrics that aid the understanding of the goodness of each model in their training phase while being general enough to allow the comparison across CL and FL implementations. Therefore, we choose the following set of metrics:
\begin{enumerate}
    \item \textit{Balanced Accuracy (BA)}, a commonly used~\cite{vaizman2018context} metric in label-imbalanced settings. It is the macro-averaged recall across all labels.
    \item \textit{Macro-averaged F1-Score}, similarily to BA, in class-imbalanced settings, it can assess the models predictive power across all classes, without the bias toward the majority class introduced by the label imbalance.
    \item \textit{F1-Score} on the \textit{minority class Running} showcases the model's capability to predict severely underrepresented classes.
    \item \textit{F1-Score} on the \textit{majority class Sitting} on the hand shows how the model works when there is enough information.
\end{enumerate}

We use these metrics for tuning, comparing and selecting the best setting for our CL and FL models. Furthermore, they work as a reference during the evaluation of the presented method, in Section~\ref{section:evaluation}.

\subsection{Model Design \& Tuning} \label{subsec:hyperparameters_tuning}
We model the problem with a multi-layer perception (MLP) with 64, and 16 neurons in each layer. Leaky ReLU was used as the activation function of the hidden layer and softmax as the output activation. The model architecture (activation functions and number of neurons) is inspired by the work performed by the Extrasensory dataset authors \cite{vaizman2018context}, allowing us to have a direct comparison. The main difference to their work is that we perform multi-class classification on the mutually exclusive labels, whereas they focus on multi-label distribution on many different subsets of labels and also experiment with different layer/neuron number combinations. As we employ softmax as the activation, we use categorical cross entropy as the loss function. SGD optimizer was used with momentum set to a common value of 0.9.

First, we focus on the CL model.
We tune the batch size and learning rate to find the optimal configuration, using the \textit{base-augmented} dataset. We evaluate the model with the following hyperparam grid: $batch\ size\ \in \{32, 64, 128, 256\}$  and $learning\ rate\ \in \{ 0.01, 0.001 \}$. We monitor the \textit{convergence} as well as the performance of the models on the \textit{test set} using the metrics defined in     Section~\ref{subsec:qam}. We display the average of the last 5 epochs for $BA$, $Macro\ F1$ and $F1\ Score$ on \textit{Running} class in Table \ref{tab:hyperparam_opt}.
Overall, the $[BS=256,\ LR=0.01]$ configuration outperforms the others on almost all the metrics. These results tell us that the CL model can accurately perform HAR, especially with the aforementioned configuration, reaching up to $\approx 0.7$ for $BA$ and $Macro\ F1$ and $\approx 0.64$ on the \textit{Running} $F1\ Score$.

As a result of multiple preliminary experiments, our configuration for synchronous FL is as follows: we train each model for a maximum of 100 rounds with 2 local epochs on all 60 clients, as the FL model requires, overall, more epochs for converging than the CL one. We varied the number of clients per round $S \in {20, 40, 60}$ and found that they all achieved comparable performance, where the larger values implied fewer fluctuations in the observed metrics between each round.  Therefore we select all $60$ clients in each round to make the results comparable to the AFL setting where all 60 clients run continuously. For lower time complexity, we use early stopping; the training process is truncated if the performance doesn't improve after 50 rounds. On top of this setup, we tune the BS and LR, using the same values as used in CL. Table \ref{tab:hyperparam_opt} contains the average of the last 5 values of hyperparameter tuning results for this setting. We can immediately notice that the overall scores for $BA$ and $Macro\ F1$ are lower than in CL ($\approx 0.6\ \text{vs.} \approx 0.7$). This degradation seems to impact more the less-represented labels, e.g., Running class ($\approx 0.5\ \text{vs.} \approx 0.6$ in CL), than the majority ones. The $F1\ Score$ for the Sitting class is in fact only slightly lower.


In asynchronous FL, for hyperparameter tuning, we train all the models for 40 minutes. The goal is to select the model with the highest F1-Score as the baseline defined by batch size and learning rate. Table \ref{tab:hyperparam_opt} summarizes the performance of the models trained with different hyperparameters, which are comparable to the synchronous FL tuning results presented in the same Table. Ultimately, we select the $[BS=128,\ LR=0.01]$ configuration as it achieves the best performance across the majority of tracked metrics. Likewise, we performed the experiments for various mixing ratio settings $\alpha_{MR} \in \{ 0.2, 0.4, 0.8 \}$. We discovered that only the speed of convergence changes with different values. Therefore, to reach maximum convergence  $\alpha_{MR} = 0.8$ was selected.

To summarize the results of hyperparameter tuning across all three paradigms of training: CL, SFL and AFL, we present the scores of each metric averaged across the last five convergence observations in Table \ref{tab:hyperparam_opt} for different batch size and learning rate configurations.

\begin{table}
\small

\begin{tabular}{lr|rrrr|rrrr}
\toprule
  & LR & \multicolumn{4}{r|}{0.001} & \multicolumn{4}{r}{0.01} \\
  & BS & 32 & 64 & 128 & 256 & 32 & 64 & 128 & 256 \\
\midrule
\multirow[t]{3}{*}{CL} & m-F1 & 0.69 & 0.70 & 0.59 & 0.58 & 0.60 & 0.68 & 0.60 &\textbf{ 0.71} \\
 & BA & 0.69 & 0.70 & 0.60 & 0.59 & 0.66 & 0.67 & 0.61 & \textbf{0.71} \\
 & F1-R & 0.54 & 0.60 & 0.00 & 0.00 & 0.45 & 0.49 & 0.00 &\textbf{ 0.64} \\
\cline{1-10}
\multirow[t]{3}{*}{SFL} & m-F1 & 0.61 & 0.59 & 0.50 & 0.43 & 0.62 & 0.62 & \textbf{0.63} & 0.61 \\
 & BA & 0.60 & 0.59 & 0.53 & 0.47 & 0.61 & 0.61 & \textbf{0.61} & 0.59 \\
 & F1-R & 0.41 & 0.43 & 0.25 & 0.00 & 0.46 & 0.52 & \textbf{0.53} & 0.49 \\
\cline{1-10}
\multirow[t]{3}{*}{AFL} & m-F1 & 0.58 & 0.56 & 0.46 & 0.44 & 0.62 & 0.62 &\textbf{ 0.63} & 0.61 \\
 & BA & 0.59 & 0.56 & 0.50 & 0.47 & 0.60 & 0.61 & \textbf{0.61} & 0.60 \\
 & F1-R & 0.36 & 0.29 & 0.00 & 0.09 & 0.48 & 0.46 & \textbf{0.54} & 0.47 \\
\bottomrule
\end{tabular}
\caption{Summary of the batch size (BS) and learning rate (LR) hyperparameter tuning for the three settings: Centralized Learning (CL), Synchronous Federated Learning (SFL) and Asynchronous Federated Learning (AFL). The monitored metrics are macro-averaged F1 score (m-F1), balanced accuracy (BA) and F1 score on the underrepresented running class (F1-R). As the performance on  F1 score on the sitting class did not significantly vary, it was omitted from the table. }
\label{tab:hyperparam_opt}
\end{table}
\vspace{-12pt}

\section{Evaluation} \label{section:evaluation}

In this section, we evaluate the challenges or complications that may arise when transitioning from CL to FL models in non-IID settings. We first describe our evaluation framework. Then, we group the design decisions (presented in Section \ref{section:methodology}) into two groups: 1) data-related decisions and 2) system/model-related decisions and describe their impact on FL model training and evaluation.

\vspace{-10pt}
\subsection{Evaluation Testbed Setup}
We need to consider various aspects when setting up the \textit{evaluation testbed}. First, we need to guarantee that the results produced for AFL are comparable with the other methods. To do so, we (1) extract performance metrics values after a fixed number of updates (instead of a fixed training duration) and (2) consider two evaluation vantage points (VP), central (2a) and distributed (2b). In the \textit{central} VP (2a), we perform the evaluation on the centralized test set on the server. In synchronous settings, this step is performed before starting the round; in AFL, we perform it periodically (every $20s$). In the \textit{distributed} VP (2b), the evaluation happens locally on each client before the local training starts. In this case, we use the local client validation set.
Finally, we need to specify the \textit{performance metrics}. As model quality indicators, we leverage the metrics defined in Section~\ref{subsec:qam}, i.e., $BA$, $Macro\ F1$, $F1\ Score$ for \textit{Running}, and $F1\ Score$ for \textit{Sitting}. 


For the \textit{execution setup}, we reproduce clients and server by running a ray simulation provided by the Flower framework~\cite{beutel2020flower}. We perform our experiments on a Ubuntu 24.04 LTS VM with the AMD EPYC 7742 64-Core CPU and 384GB of memory.

\subsection{Data-related Decisions}

We now evaluate the effects of data preprocessing decisions and which pitfalls they might lead to when transitioning from CL to FL in non-IID settings.


\subsubsection{Fair vs Hold-out Test set} \label{subsec:test_scheme}

There are two ways for generating the test set in federated learning settings with IoT data, i.e., with multiple clients: (1) with \textbf{Hold-Out Clients} (HOC), that keeps only the data of a \textit{subset of clients} for testing, or (2) by using a \textbf{Fair Test Set} (FTS), where \textit{each client selects a portion} of data to send to the server to create the test set.
HOC (1) offers better privacy guarantees, but the final test set suffers from client-selection bias. This effect is even more prominent in severe non-IID settings, amplifying quantity, feature, and label skews. FTS (2) is weaker from the privacy perspective but enables a fairer evaluation of the global model performance by sampling data from all clients.

Our work targets the fair examination of the models; therefore, we use FTS. Especially considering our use case, the skew in label distribution is so accentuated that with HOC, we could even leave out some categories altogether, leading to a partial evaluation.


\takeaway{The client-selection bias present in HOC is amplified by the non-IIDness of typical IoT data. FTS can be used to gain a clear and stable view on the global data distribution enabling fair evaluation while carrying privacy risks.}


\subsubsection{Data augmentation and its effects in non-IID FL} \label{subsec:data_augmentation}

\begin{figure}
    \centering
    \includegraphics[width=1\linewidth]{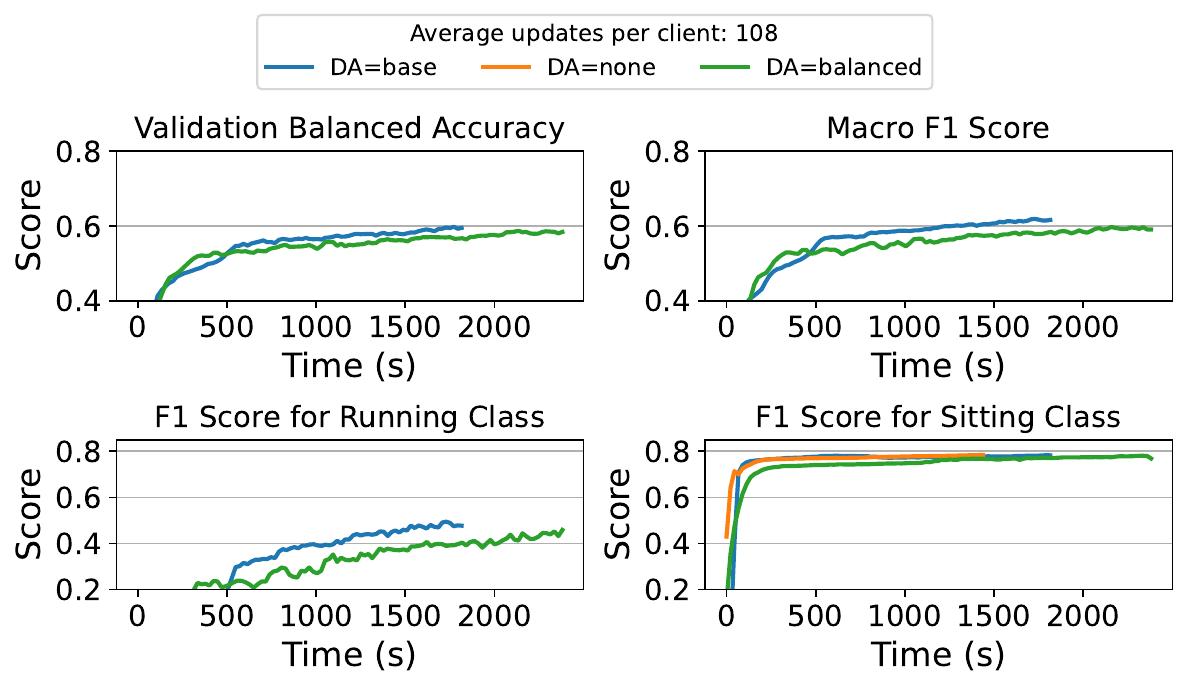}
     \captionsetup{skip=-2pt}
    \caption{Zoomed-in convergence plots of the evaluation metrics for different data augmentation schemes (DA) in asynchronous FL. Different line length is the result of fixing a number of average client updates rather than train time.}
    \label{fig:varying_data_augmentation_async}
\end{figure}

Here we re-visit these augmentation schemes from Section \ref{subsubsec:data_augmentation} only within the \textit{federated learning context}. Figure~\ref{fig:varying_data_augmentation_async} depicts the convergence results of training the AFL baseline model with varied data augmentation. Likewise, a summary of the results for CL, SFL, and AFL is presented in Table \ref{tab:evaluation}. Across all three settings, the \textit{running} label is almost always ignored by the models. The model trained with no augmentation performs comparably on the majority label as the one trained with base augmentation. However, excessive augmentation can be problematic. The \textit{balanced} setting, despite equal class representation, introduces oversampling bias, making it better than no augmentation but worse than moderate augmentation.

\takeaway{Adding Gaussian noise as data augmentation scheme partially addresses the label skew in CL, SFL, and AFL. The magnitude of the performed data augmentation is a tunable parameter as excessive augmentation can lead to oversampling bias and no augmentation leaves the minority classes severely underrepresented and ignored by the model.}

\subsubsection{Global Data Scaling and Persistence of Feature Skew } 
\label{subsec:scaling}

\begin{figure}
    \centering
    \includegraphics[width=1\linewidth]{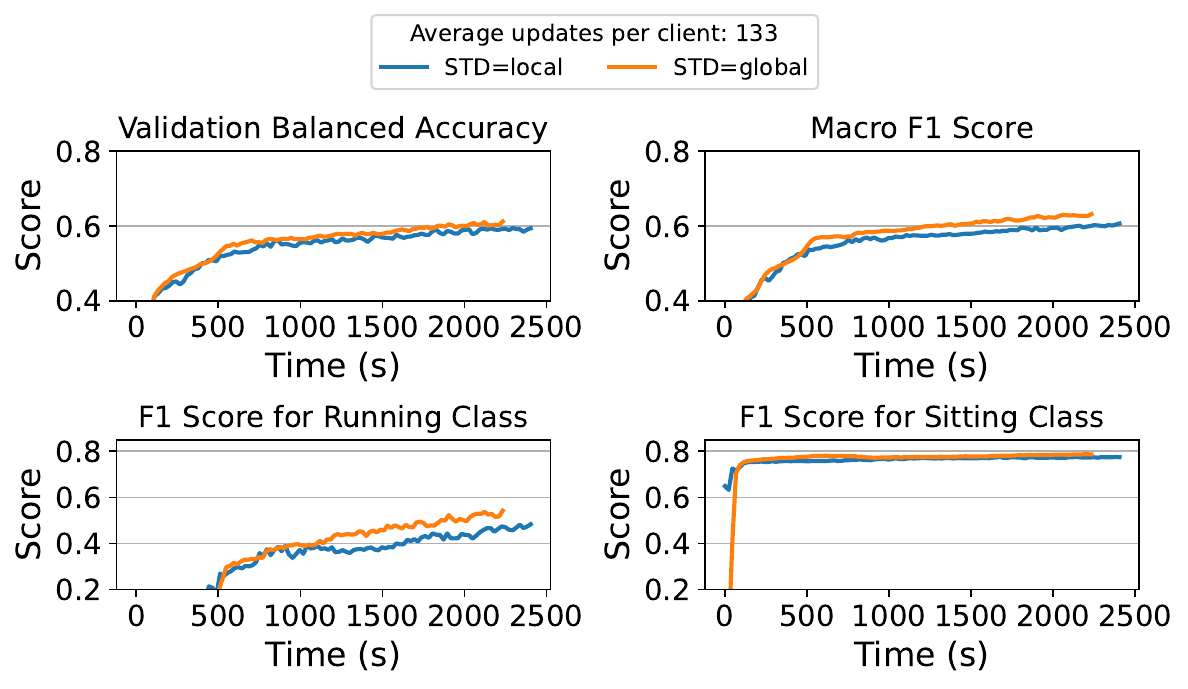}
    \captionsetup{skip=-2pt}
    \caption{Zoomed-in convergence plots of the evaluation metrics for different feature standardization schemes (STD) in asynchronous FL.}
    \label{fig:global_vs_local_std_async}
\end{figure}

We further evaluate the influence of two data scaling (standardization) methods on model convergence in asynchronous FL. In the first approach, we scale the features of the data based on \textbf{client-local} means and standard deviations. In the second one, we first share the local means and standard deviations with the server before training, and then we scale the data on all clients with the aggregated \textbf{global} mean and standard deviation. 
While carrying privacy risks due to the communicated statistics, the second, \textit{global}, method improves the performance of the models in all four tracked metrics, as it is visible from the Figure \ref{fig:global_vs_local_std_async} and Table \ref{tab:evaluation}. Individuals' personal habits and way of performing them usually influence their local feature distributions  (e.g. every individual cycles differently and introduces different sensor readings for this class). Secondly, additional feature skew among the clients stems from \textit{device heterogeneity}. As a result, scaling the features \textit{locally} tends to keep a considerable proportion of the already present feature skew and provides skewed data views for the model degrading its performance.

\begin{figure}
    \centering
    \includegraphics[width=0.9\linewidth]{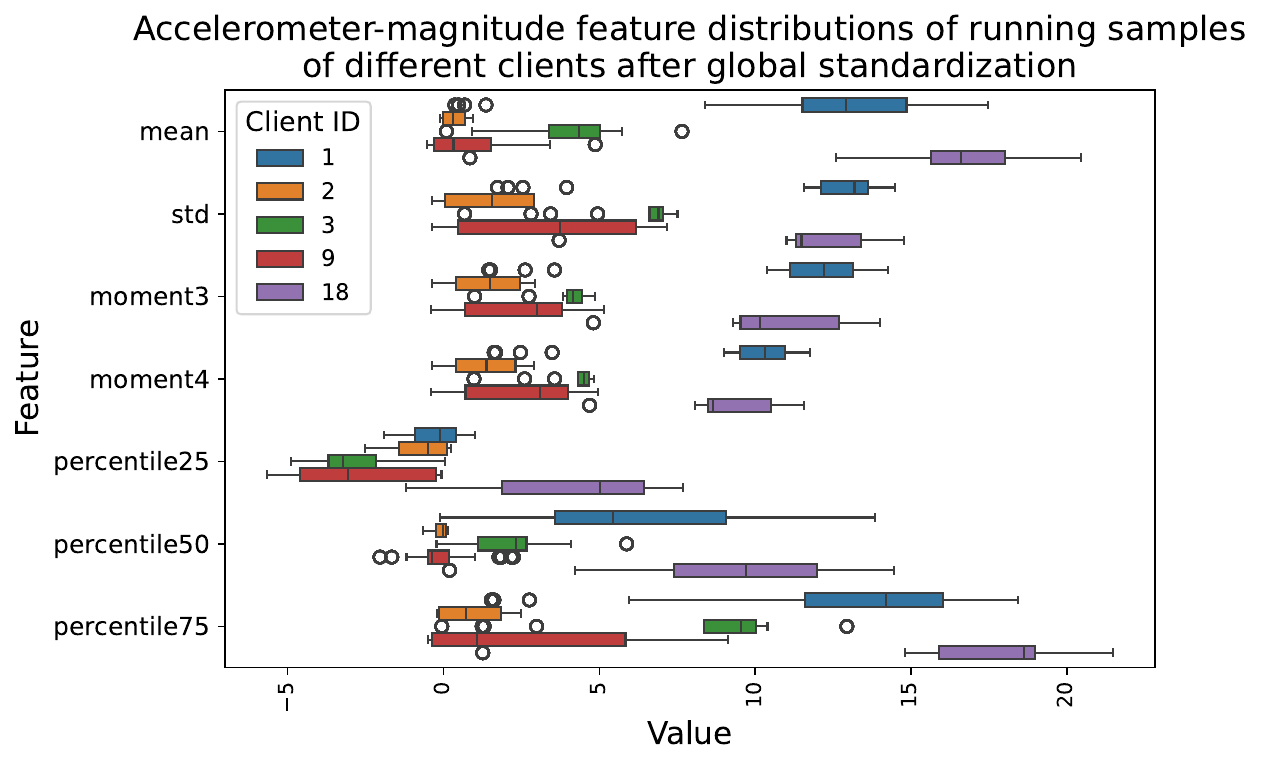}
     \captionsetup{skip=-2pt}
    \caption{Boxplots depicting features distributions of the \textit{running} class samples for different clients (varied by color) [Sensor: accelerometer-magnitude]}
    \label{fig:running_feature_skew}
\end{figure}

\textit{Global} scaling addresses these causes for feature skew, however, a significant amount of feature skew remained for the running class samples as depicted in Figure \ref{fig:running_feature_skew}. This skew is potentially a result of a difference in local label distributions. As each client has a different ratio of the samples of each class, the feature means and standard deviations will be skewed towards each client's distribution \textit{as well}. As the minority class usually has fewer samples on each client, the effects of such skews are more prominent in minority classes. 

\takeaway{Scaling the data globally, significantly improves the convergence of the models, but it introduces privacy risks. Even with global data standardization, the clients might still have different representations of the same class. This proves that feature skew is a persistent issue in non-IID IoT (HAR) datasets and is amplified in the minority classes.}

\subsection{System \& Model-related Decisions}
\label{subsec:adam_vs_sgd}

In the following, we focus on the effect of different optimizers on model performance with non-IID data and the implications of server-processing delays on AFL with non-IID data.





\begin{figure}
    \centering
    \includegraphics[width=1\linewidth]{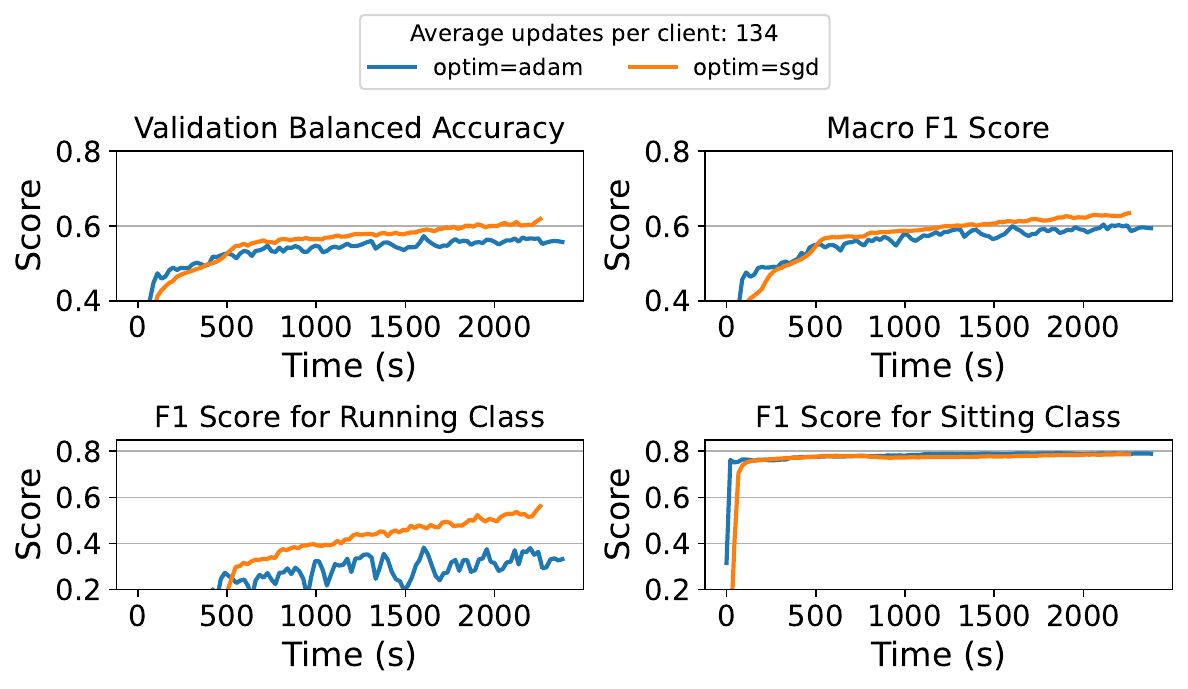}
     \captionsetup{skip=-2pt}
    \caption{Zoomed-in convergence plots of the evaluation metrics for different optimizer (optim) in asynchronous FL.}
    \label{fig:adam_vs_sgd_async_fl}
\end{figure}

\subsubsection{Optimizer Selection}
Here, we compare two optimizers, ADAM~\cite{kingma2014adam}, commonly used in FL settings, and SGD with momentum (SGD-m), a performative optimizer in Deep Learning. To examine the magnitude of the optimizer's impact on the model performance, we fix all the other training parameters. 
We present the results in Figure~\ref{fig:adam_vs_sgd_async_fl} and Table \ref{tab:evaluation}.  ADAM converges faster but with lower overall performance. Moreover, Figure~\ref{fig:adam_vs_sgd_async_fl} depicts how training using ADAM has a much more unstable behavior, obviously shown in the case of \textit{Running} (bottom left)).
We assume that the unsatisfactory results with the ADAM optimizer are due to extremely short local train times (2 epochs only), leading to restarting the ADAM state with each round of training and essentially forgetting it. At the same time, leveraging na\"ive approaches such as extending the local train time or locally maintaining the ADAM state are not recommended. First, significantly increasing the (local) train time would lead, in FL settings, to local gradients that diverge toward the clients' local distributions, causing issues during aggregation. On the other hand, maintaining the local ADAM state and keeping fewer local epochs carries the risk of ADAM overfitting to the local data.
To make the momentum and learning rate adaptive and utilize ADAM's full potential, one could consider either applying ADAM on the server side, as presented in \cite{leroy2019federated, ju2023accelerating}, or sharing ADAM state among the clients, as presented in \cite{karimireddy2020mime}. Exploring these solutions is however outside of the scope of this paper.

\takeaway{Applying SGD-m as the optimizer in FL settings has proven to be more advantageous than applying ADAM. As ADAM contains more stateful parameters that are tracked over multiple epochs, the approach does not perform well in, typically stateless, federated optimization. To improve the performance of ADAM, one can either apply ADAM on the server or share the optimizer's state with the server during the entire training.}

\subsubsection{Effects of Busy Servers} 
\label{subsec:server_delay}

\begin{figure}
    \centering
    \includegraphics[width=1\linewidth]{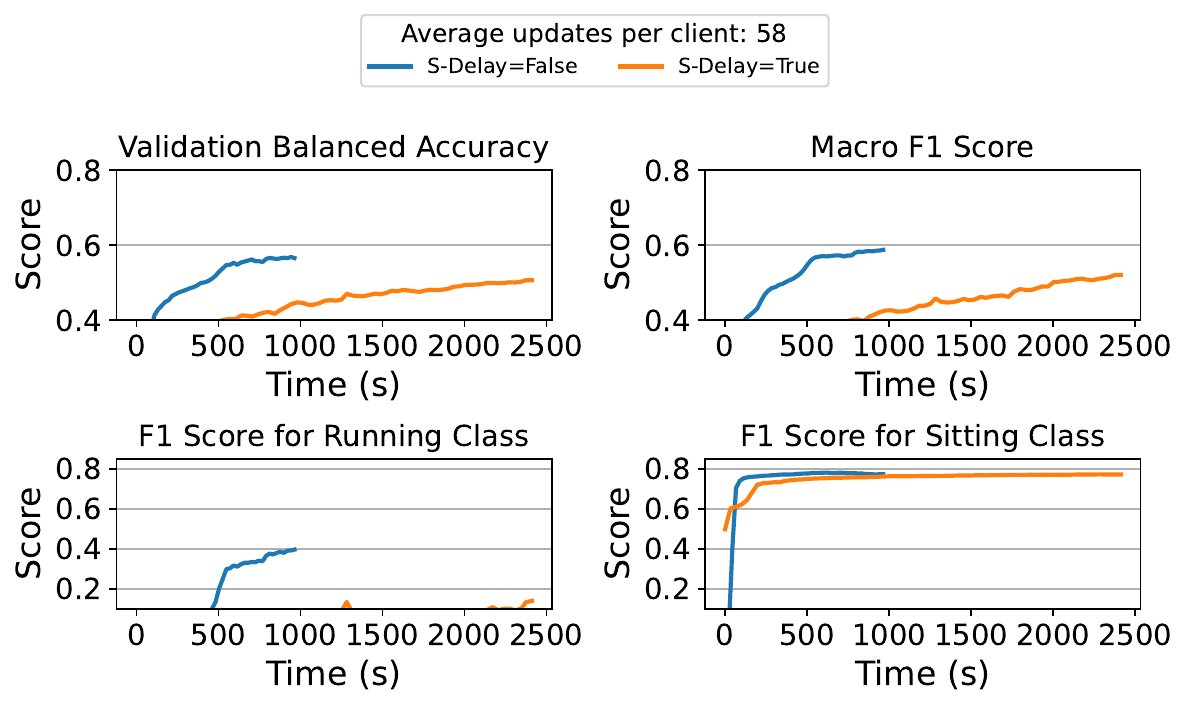}
     \captionsetup{skip=-2pt}
    \caption{Zoomed-in convergence plots of the evaluation metrics for different server-delay (S-Delay) amounts (True if additional delay was introduced.) in asynchronous FL.}
    \label{fig:async_server_delay}
\end{figure}

Custom FL workflows typically include additional processing on the server (e.g. for clustering \cite{ge2024fedaga, briggs2020federated}) that introduces delays. To examine the effect of these delays on AFL with non-IID data we simulate them by adding busy waiting on the server on two vantage points (VPs): (1) before centralized evaluation we add a busy wait of $10$ seconds and (2) before each client's update is merged into the global model we add $1$ second of busy wait. The results of this run compared to the baseline (without the delays) are illustrated in Figure \ref{fig:async_server_delay}. The model trained without the delays naturally reaches the fixed number of updates sooner (hence the shorter line). However, adding delay to the server significantly degrades model performance even with the \textit{same} number of average updates per client. This discrepancy is especially visible in the minority class \textit{running}. This observation highlights the importance of timely model updates and underscores the need for adequate server resources and time-efficient adaptation techniques, even without direct model training expected on the server.

\takeaway{For AFL workflows involving intensive computations, achieving the baseline without additional computations may be difficult due to the performance discrepancy caused by server delays.}




\begin{table}[]
    \small
    \begin{tabular}{llrrrrrrr}
    \toprule
     & Param & \multicolumn{3}{r}{data augmentation} & \multicolumn{2}{r}{scaling} & \multicolumn{2}{r}{optimizer} \\
     & Value & none & base & bal & glob. & loc. & adam & sgd-m  \\
    \midrule
    \multirow[t]{3}{*}{CL} & m-F1 & 0.47 & 0.71 & 0.68 & - & - & - & -  \\
     & BA & 0.46 & 0.71 & 0.69 & - & - & - & -  \\
     & F1-R & 0.00 & 0.64 & 0.60 & - & - & - & -  \\
    \cline{1-9}
    \multirow[t]{3}{*}{SFL} & m-F1 & 0.51 & 0.63 & 0.61 & 0.63 & 0.61 & 0.59 & 0.63 \\
     & BA & 0.47 & 0.61 & 0.59 & 0.61 & 0.61 & 0.56 & 0.61  \\
     & F1-R & 0.00 & 0.53 & 0.49 & 0.53 & 0.44 & 0.31 & 0.53 \\
    \cline{1-9}
    \multirow[t]{3}{*}{AFL} & m-F1 & 0.39 & 0.62 & 0.59 & 0.63 & 0.60 & 0.59 & 0.63 \\
     & BA & 0.39 & 0.60 & 0.58 & 0.60 & 0.59 & 0.56 & 0.61  \\
     & F1-R & 0.00 & 0.48 & 0.45 & 0.53 & 0.48 & 0.32 & 0.53  \\
    \cline{1-9}
    \bottomrule
    \end{tabular}
     \captionsetup{skip=-10pt}
    \caption{Summary of the influence of \textit{data augmentation}: none, base and balanced (bal), \textit{standardization (scaling)} : local and global, and \textit{optimizer}: SGD-m and ADAM on model performance. Only the federated settings were considered for the effects of data \textit{scaling} and \textit{optimizer}. As the performance on  F1 score on the sitting class did not significantly vary, it was omitted from the table. }
    \label{tab:evaluation}
\end{table}

\section{Related Work}
\label{sec:related_work}
We categorize previous contributions based on the two main topics of our research. First, we focus on related work on FL applied to Human Activity Recognition. Furthermore, we present the findings of other authors on the issues that non-IID data brings to FL. 

\textit{Human Activity Recognition with Deep and Federated Learning}

Better devices' hardware, together with higher privacy awareness, led HAR research to focus on using FL approaches. Sozinov et al.~\cite{sozinov2018human} train a deep neural network to model the Heterogenous HAR data.
They prove that the model performance in both balanced and simulated imbalanced settings is acceptable for this dataset. However, the applicability of their insights to the Extrasensory dataset~ \cite{stisen2015smart} is limited as the actions performed have a uniform label distribution. Smith et al.~\cite{smith2017federated} introduce MOCHA, a multi-task learning framework for addressing both statistical as well as system-level heterogeneities in FL, which was evaluated on the UCI-HAR dataset \cite{anguita2013public}. 
More specifically, for non-IID data in asynchronous FL, an approach~\cite{chen2020asynchronous, li2020federated_fedprox} is to learn shared feature representations on the server, together with a decayed update coefficient. The goal is to the previous global model with new updates and improve the local loss function. Chen et al.~\cite{chen2021asynchronous} integrate this method with a drift detection and drift correction scheme to address and test it on the Extrasensory dataset.
However, these contributions solely use the HAR dataset as an evaluation. \textbf{Our work} instead proposes a fully IoT/HAR-oriented analysis approach. We offer a thorough examination of the HAR use case from both IoT and FL perspectives. We find that the typical feature learning approaches usually work around the issue of non-IID data without trying to understand it fully. In contrast, we aim to provide an in-depth analysis of the issues and causes of performance degradation.

\textit{Issues with Non-IID data in Federated Learning} 
There are two primary surveys addressing the topic of non-IID data in FL \cite{zhu2021federated} and \cite{lu2024federated}. The former lays a basis for the classification of approaches to non-IID data in FL and identifies three groups of solutions (\textit{data-, algorithm-} and \textit{system-oriented} solutions), while the latter focuses on the evolution of FL research that happened in the years after the publishing of the former survey. While these surveys offer a broad overview of the non-IID data issue, they do not analyze the influence of different hyperparameters, model evaluation schemes or different data scaling methods. Contrary to their work, we focus on the method of transitioning from centralized to federated learning and the effects of various design decisions on model performance and model evaluation properties.

\textit{Transitioning from Centralized to Federated Learning} 
The most similar contribution to our work is presented by Drainakis et al.\cite{drainakis2023centralized}. They evaluate the effects of transitioning from centralized (CL) to FL settings in non-IID data scenarios. In contrast to our research, their focus is on the network resources and energy consumption. Furthermore, while they compare CL and FL performance, they do not work towards mitigating non-IIDness in FL and do not consider AFL setting.

\section{Conclusion}
\label{sec:conclusion}

In this paper, we presented a 
methodology for transitioning from CL to SFL/AFL in the IoT landscape with non-IID settings. We discussed the critical design decisions of the IoT FL model development lifecycle. 
We illustrated our methodology on a realistic HAR case study to analyze its practical feasibility in real-world settings.  
%
We evaluated the implications of data- and model-related design decisions on the model development in IoT FL systems. Based on our evaluation results we identified the following main takeaways to support the researchers and practitioners developing FL IoT systems:  
1) generating a stable and fair test set carries privacy risks but is crucial to ensure reliable performance,
2) data augmentation is a tunable parameter that can significantly improve performance in non-IID FL, 
3) global data scaling carries privacy risks while offering a more consistent view of the data, 
4) even after global standardization feature skew often remains (especially in minority classes),
5) using a state-based optimizer such as ADAM degrades the performance in non-IID FL 
and 6) additional delays on the server degrade performance and influence comparability.
To obtain the AFL results presented in this work we implemented an open-source extension of the Flower framework which supports AFL.

\textit{Limitations and Potential Concerns}
While the use of a single dataset may raise concerns about the generalizability of the findings, the \textit{Extrasensory} dataset is an excellent representative of a realistic and naturally non-IID HAR setting. Therefore, the takeaways of our exhaustive experimentation with this dataset create a stable blueprint for similar tasks and datasets across the IoT and HAR fields. While global scaling might introduce privacy risks, we point it out as one of the trade-offs, i.e. stronger privacy vs better performance. Further discussion on the severity of this privacy risk and privacy-preserving techniques that could be used to mitigate this issue is out of the scope of this work. Similarly, there are more sophisticated techniques for addressing the persistent feature skew such as an adaptive learning rate or weighted loss functions for minority classes, however, they typically require additional assumptions and their effectiveness may vary depending on the use case. On the contrary, our experiments were guided by the typical ML pre-processing steps that can be applied immediately, and therefore, a discussion of these advanced methods is out of the scope of this work. 

\textit{Future work}
In the future, we intend to continue our work in several directions. After establishing the performance degradation present in non-IID federated HAR settings, we aim to further investigate the model update gradients and develop an adaptive asynchronous solution that leverages this approach. This solution will have the goal of improving the performance of underrepresented classes and, by proxy, the general model performance. We also intend to extend our current solution to streaming and continuous learning settings which are important for the FL systems in IoT settings. Our further aim is to advance our approach from the system perspective, enabling the provisioning and governance~\cite{nastic2015governinguncertainty} of FL models in the IoT and the Edge-Cloud landscape.






\begin{acks}
This work is partially funded by the Austrian
Research Promotion Agency (FFG), under the project No.
903884.
The work of Anastasiya Danilenka was conducted during the research visit funded by the Warsaw University of Technology within the Excellence Initiative: Research University (IDUB) programme. The work of Andrea Morichetta is funded by the HORIZON Research and Innovation Action 101135576 INTEND ``Intent-based data operation in the computing continuum.''
\end{acks}

\bibliographystyle{ACM-Reference-Format}
\bibliography{refs}


\end{document}